# Normative brain mapping using scalp EEG and potential clinical application

Vytene Janiukstyte[1], Thomas W Owen[1], Umair J Chaudhary[3], Beate Diehl[3], Louis Lemieux[3], John S Duncan[3], Jane de Tisi[3], Yujiang Wang[1,2,3], Peter N Taylor[1,2,3,*]

1. CNNP Lab (www.cnnp-lab.com), Interdisciplinary Computing and Complex BioSystems Group, School of Computing, Newcastle University, Newcastle upon Tyne, United Kingdom, NE4 5DG

2. Faculty of Medical Sciences, Newcastle University, Newcastle upon Tyne, United Kingdom, NE2 4HH

3. Department of Clinical and Experimental Epilepsy, UCL Queen Square Institute of Neurology, University College London, Queen Square, London, United Kingdom, WC1N 3BG

## Abstract

A normative electrographic activity map could be a powerful resource to understand normal brain function and identify abnormal activity. Here, we present a normative brain map using scalp EEG in terms of relative band power. In this exploratory study we investigate its temporal stability, its similarity to other imaging modalities, and explore a potential clinical application.

We constructed scalp EEG normative maps of brain dynamics from 17 healthy controls using source-localised resting-state scalp recordings. We then correlated these maps with those acquired from MEG and intracranial EEG to investigate their similarity. Lastly, we use the normative maps to lateralise abnormal regions in epilepsy.

Spatial patterns of band powers were broadly consistent with previous literature and stable across recordings. Scalp EEG normative maps were most similar to other modalities in the alpha band, and relatively similar across most bands. Towards a clinical application in epilepsy, we found abnormal temporal regions ipsilateral to the epileptogenic hemisphere.

Scalp EEG relative band power normative maps are spatially stable across time, in keeping with MEG and intracranial EEG results. Normative mapping is feasible and may be potentially clinically useful in epilepsy. Future studies with larger sample sizes and high-density EEG are now required for validation.

## Introduction

Abnormal electrographic brain activity can manifest clearly in a wide range of neurological disorders such as Alzheimer's Diseases, Epilepsy, Parkinson's Diseases and more (Bonanni et al., 2008; Bosch-Bayard, Aubert-Vazquez, et al., 2020; Czigler et al., 2008; Owen et al., 2022; Sun et al., 2014; Taylor et al., 2022; Varatharajah et al., 2021). For example, in epilepsy, spikes and sharp waves are commonly observed interictally (Abel et al., 2018; Hauser et al., 1982; Kaiboriboon et al., 2012; Varatharajah et al., 2021; Wennberg & Cheyne, 2014). These features can often be identified visually; however, extracting subtler changes in cortical dynamics is much more challenging.

Subtle changes to electrographic activity may be difficult to detect in patients because heterogeneity in regional bandpower exists even in recordings from healthy subjects. For example, alpha oscillations are much stronger in parietal and occipital than in frontal or temporal brain regions (Frauscher et al., 2018; Groppe et al., 2013; Nayak & Anilkumar, 2022). Theta activity is heightened by drowsiness and some cognitive tasks in frontal midline areas (Groppe et al., 2013; Nayak & Anilkumar, 2022; Niso et al., 2019; Snipes et al., 2022). Other spatial profiles include delta activity in orbitofrontal areas and in the temporal lobe (Frauscher et al., 2018; Niso et al., 2019), and beta oscillations dominating the motor areas (Barone & Rossiter, 2021; Morillon et al., 2019). Any excess or deficit in power, in a particular frequency at an unconventional cortical location may indicate pathological activity (Bonanni et al., 2008; Bosch-Bayard, Aubert-Vazquez, et al., 2020; Czigler et al., 2008; Sun et al., 2014; Varatharajah et al., 2021), and can even be affected by medication (Jensen et al., 2005). Thus, mapping the range of healthy brain dynamics may help to identify subtle pathological events, which often appear concealed by the healthy spatial variations.

Previously, the Brain Electrical Activity Mapping (BEAM) method was proposed as a potential diagnostic tool to identify underlying abnormalities (Duffy et al., 1979; Duffy et al., 1989). BEAM maps condense scalp EEG data into a topographic map to show mean spectral energy across any individual frequency band, in an individual subject. More recently, quantitative EEG (QEEG) was utilised to derive subtle features or specific patterns. The role of QEEG has been widely explored in the literature as a promising tool for novel discoveries and biomarkers in various brain diseases and syndromes (Arns et al., 2013; He et al., 2017; Ko et al., 2021; Livinț Popa et al., 2021; Tedrus et al., 2019). There is a view that the lack of standardization across QEEG databases is a limitation. Recently, standardized normative EEG databases accounting for the effects of age and sex have been published (Bosch-Bayard, Aubert-Vazquez, et al., 2020; Keizer, 2021; Ko et al., 2021; Valdes-Sosa et al., 2021). Here, we propose an extension of quantitative EEG brain mapping by extracting frequency band power across healthy controls and averaging individual band powers to construct relative power normative maps.

Comparing patients with normalised healthy controls to expose pathology is common in neuroimaging, and is gaining traction in neurophysiology. In recent studies, neurophysiology data from MEG and intracranial EEG (iEEG) have been used successfully to identify abnormalities (Bernabei et al., 2022; Bosch-Bayard, Aubert-Vazquez, et al., 2020; Owen et al., 2022; Taylor et al., 2022). However, MEG is currently expensive in many circumstances, and iEEG is highly invasive and costly (Arya et al., 2013; Blauwblomme et al., 2011). Scalp EEG has therefore been reconsidered for its inexpensive and non-invasive advantages (Arya et al., 2013; Bosch-Bayard, Aubert-Vazquez, et al., 2020; Bosch-Bayard, Galan, et al., 2020; Fitzgerald et al., 2021; Michel & Brunet, 2019). Despite these advantages and previous attempts to utilise scalp EEG as a diagnostic tool, scalp EEG normative maps are not typically used in clinical routine investigations.

In this exploratory study, we construct normative maps across the five classical frequency bands using scalp EEG from 17 healthy subjects. Initially, we assess and illustrate the consistency of the normative maps across multiple segments of data. Next, we relate the normative band power distributions to those derived from MEG and intracranial EEG to investigate their similarity. Finally, using a temporal lobe epilepsy cohort of 22 patients, we illustrate a potential clinical application of scalp EEG normative mapping to lateralise abnormal temporal regions across the cohort.

## Methods

### Scalp EEG subjects

The scalp EEG data were acquired from 17 healthy volunteers and from 22 patients with temporal lobe epilepsy (TLE) undergoing presurgical evaluation. The selection criteria were: subsequently underwent curative surgery for TLE and their resting-state scalp recording contained at least 60 seconds of artefact-free EEG. The patients had not undergone neurosurgery prior to the EEG recordings, which were performed for research purposes during pre-surgical evaluation. A summary of our cohort is provided in table 1 and further details provided in Supplementary Material 1.

Table 1: Descriptive statistics of the healthy controls and patient data

|  | Controls (17) | Patients (22) | Test statistic |
|---|---|---|---|
| Age (ys, mean, SD) | 31.9,6.46 | 34.2,10.1 | p=0.41, es=8.69 |
| Sex (M/F) (%) | 11/6 (65%) | 9/13 (41%) | p=0.14, $\chi^2$=2.17 |
| Epilepsy lateralisation (L/R) (%) | N/A | 14/8 (64%) | N/A |
| Age of epilepsy onset (yr; mean, SD) | N/A | 12.9, 8.3 | N/A |
| Duration of epilepsy (yr; mean, SD) | N/A | 21.3, 13.2 | N/A |

### EEG acquisition and processing

Eyes-closed resting-state scalp EEG data were recorded using a commercial MR-compatible system (BrainAmp MR and Vision Analyzer), using the 10-20 international system (Klem et al., 1999). The data were recorded at a sampling rate of 5000Hz, with 30-channel scalp electrodes and a common average reference. Two additional electrodes recorded EOG and ECG, capturing ocular and cardiac activity respectively. The first 30 seconds of the recordings were not considered on account of the subjects having time to settle.

Scalp EEG recordings were pre-processed using MATLAB and Brainstorm (Tadel et al., 2011) in the following steps. First, data were downsampled to 250Hz. We used ICBM152 anatomical MRI template in standard space to compute a realistic template head model, using the boundary element method (BEM). Next, we used digitized electrode locations provided by Brainstorm software and applied these to the anatomical template. Sensor time series were bandpass filtered between 0.5-47.5Hz (Fig 1a). We limited the analysis to frequencies below 47.5 Hz to for two reasons: signal degradation due to the effects of the electrical properties of skin, tissue and bone (Ramon et al., 2009; Shibata & Katsuhiro, 2018; Srinivasan et al., 1998) and mains (50Hz) artefacts. Automated cardiac detection used the ECG channel to identify standard ECG heartbeat artefact with manual intervention where necessary. Signal Space Projection (SSP) was performed to identify artefactual components, which were manually removed. The latter detection system was less successful at accurately identifying ocular artefacts. Therefore, we did not remove ocular artefacts manually to preserve biological signal, particularly given that recordings were acquired with eyes closed. Additionally, some subjects had channels removed due to other artefacts (Supplementary Material 1).

Next, we used the standardized low-resolution brain electrographic tomography (sLORETA) method and a realistic head model derived using BEM to reconstruct the EEG in source space. This created 15,000 constrained sources on the cortical surface, which were downsampled based on Lausanne parcellation that includes a variety of ROI resolutions (Hagmann et al., 2008) (Fig 1b). We use the resolution with 114 neocortical regions in the main analysis, and show others in supplementary materials. Note that current activity associated with a ROI may originate from multiple neighbouring locations, especially in lower resolution parcellations where some ROIs represent a large area of the cortex. Overlapping current magnitudes from opposite sides of the sulci are orientated in opposite directions by the constrained manifestation of source mapping. To address these issues, the source recordings were sign-flipped and averaged to produce a single time series per region across all regions size. Note that the Lausanne parcellation used in the main text is asymmetric. Therefore, to investigate asymmetries we removed six of the 114 regions in this analysis alone.

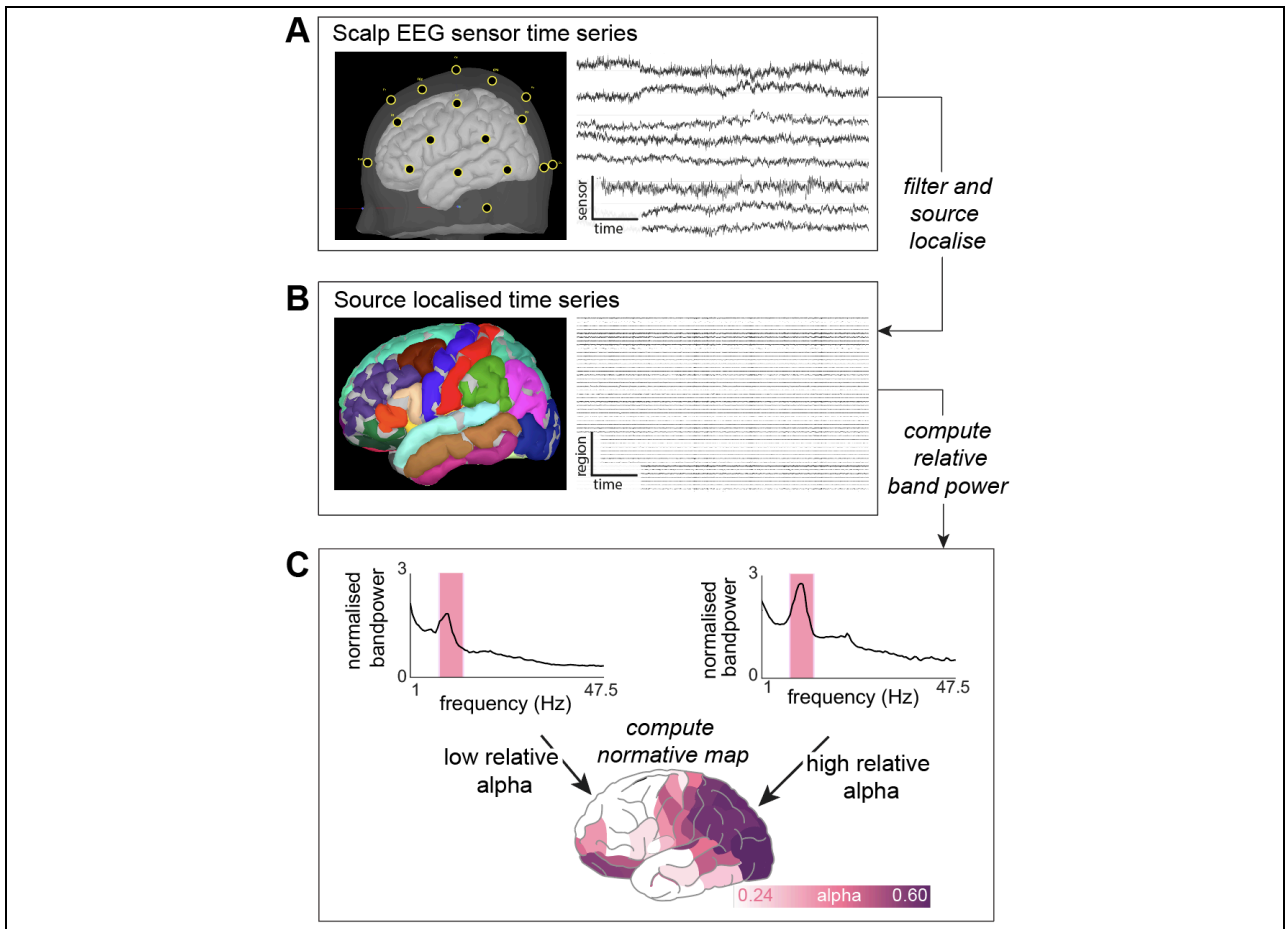

**Fig1: Scalp EEG processing pipeline.** *(A) Digitised electrode locations are projected on a cortical surface for visual verification and raw time series from 8 selected electrodes (n.b. 8 are shown for visualisation only, all are used for processing). (B) Data is filtered and source localised MRI in ICBM152 standard space, followed by parcellation. The resulting time series in source space are imported into MATLAB to extract a 60-second epoch. Source space recordings are computed to allow comparison to other modalities (iEEG,*

MEG). (C) The absolute power is normalised to compute relative power of five frequency bands across different cortical regions.

## Generating normative maps

Source time series output were imported into MATLAB and one 60-second continuous epoch was selected for each recording for further analysis. The power spectral density in each neocortical region was computed using Welch's method with a 2-second sliding window and 50% overlap for five frequency bands; delta 1–4, theta 4–8, alpha 8–13, beta 13–30 and gamma 30–47.5Hz (see Fig 1b). We scaled relative band power contributions by the total power within each band (Fig 1c).

To investigate the robustness of the normative maps to the choice of epoch, we constructed three new normative maps using 30-second non-overlapping epochs; robustness was evaluated by plotting the mean and standard deviations of regional band powers and quantified using the Spearman correlation, (Fig 2c, Fig 2d). Additional epoch comparisons can be found in Supplementary Material 2.

## Lateralising a quantified abnormality

As a proof-of-principle investigation, we next propose a potential clinical application of scalp EEG normative mapping to identify abnormalities in a temporal lobe epilepsy cohort. To this end we computed a measure of patient band-power abnormality based on the maximum band power within each ROI in the form of band power $z$-scores relative to the normative (healthy controls) spectral map using the equation:

$$\mid z_{i,j} \mid = \left| \frac{x_{i,j} - \mu_{i,j}}{\sigma_{i,j}} \right| \tag{1}$$

where $x_{i,j}$ is the patient band power and $\mu_{i,j}$, $\sigma_{i,j}$ are the healthy control band power mean and standard deviations, respectively, for ROI $i$ and frequency band $j$. As there was no prior frequency band-related hypothesis, for any given ROI we used the maximum abnormality across bands as the abnormality indicator. For the purpose of this demonstration, we focused our analysis on the 26 temporal lobe ROI of the Lausanne scale60 atlas parcellation, to test the hypothesis that abnormal band power is ipsilateral to the epileptogenic zone. We plotted the abnormality indicator (ROI maximum z-score across bands) on a representation of the temporal lobe.

## MEG and iEEG subjects

We compared our scalp EEG normative maps to maps derived from iEEG and also from magnetoencephalography (MEG) data. Eyes-closed resting-state MEG recordings were acquired for 70 healthy controls, and were processed as described previously (Owen et al., 2022). Specifically, we extracted MEG normative data between 0.5 – 47.5Hz (rather than to 77.5Hz as in Owen et al., (2023)) to match the frequency range of our scalp EEG data. In brief, relative band power spatial maps were computed using source localised MEG recordings to neocortex of healthy controls.

No iEEG recordings from healthy subjects were available. Instead, we used data from areas outside of the seizure onset zone in patients using SEEG and grid electrodes as described by (Taylor et al., 2022). In brief, electrodes were localised to cortical regions according to the parcellations described above from interictal intracranial recordings from 234 participants and the band powers extracted. Again, only power between 0.5 and 47.5Hz was analysed in the present study. Note that, normative data was collected from different subjects in each of the three modalities; MEG, scalp and intracranial EEG.

## Data & code availability

Scalp EEG normative maps in four parcellations, including cortical region lists and analysis code will be made available by the authors upon acceptance of the manuscript.

## Results

The normative distribution of power across cortical regions for each frequency band in the healthy subjects is shown in Fig 2a. We note relatively elevated delta power in most-anterior temporal and frontal regions, while alpha was prominent in the parietal and occipital regions. Furthermore, there is good right-left symmetry across frequency bands (Fig 2b). Other normative parcellation profiles show similar patterns (Supplementary Material 3).

### Normative map estimation robustness

The normative maps for different choices of epoch were highly correlated across ROI and bands, for the means and standard deviations (see Fig 2 parts c and d for epoch 1 and 2 comparison; and Supplementary Material 2 for epoch 3.)

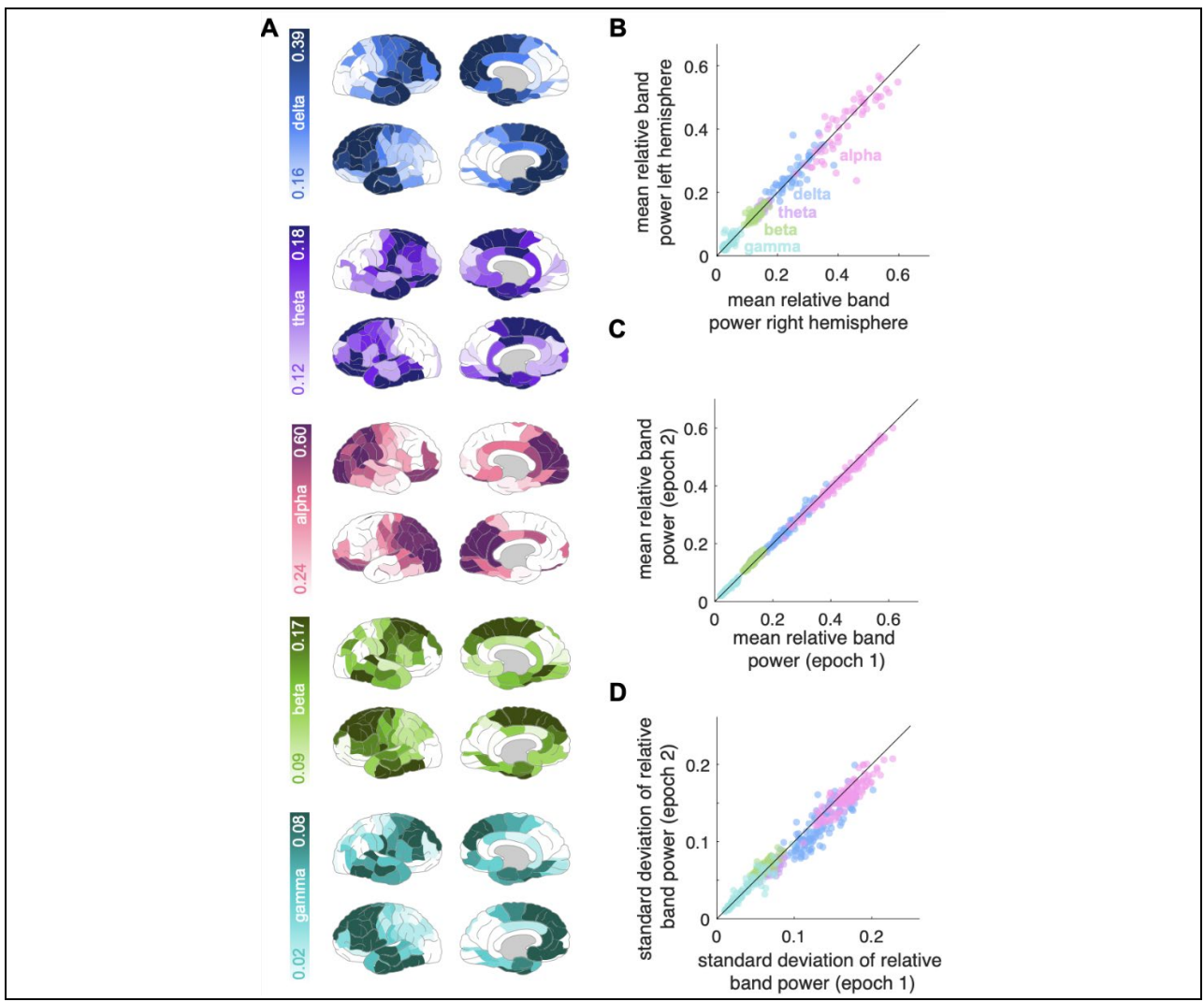

*Figure 2: Normative band power variation across regions and time. (A) Relative ROI band power averaged across the healthy subjects for the five individual frequency bands (from top to bottom: delta, theta, alpha, beta, gamma). The power density colour scales are normalised to each band's power range. (B) Left vs. right mean relative ROI band power correlation, across the five bands (same colour scheme as (A)), (rho=0.98). Each data point represents a ROI mean relative band power. Each region is therefore represented five times with different colours representing different frequencies. The identity line is plotted in black. (C-D) Reproducibility analysis. ROI band power calculated for two, 30-second, non-overlapping epochs: mean relative band power (C) (rho=0.997), and standard deviation (D) (rho=0.97).*

## Comparison of Scalp EEG, iEEG and MEG normative maps

Comparison of the EEG-derived normative maps with those generated from iEEG and MEG showed positive correlation values for all comparisons except one (Fig 3). The greatest (positive) correlations between scalp EEG and the other modalities were observed in the delta and alpha bands. In the gamma band, the scalp EEG map was positively correlated with MEG (rho=0.27), but negatively with iEEG (rho=-0.42).

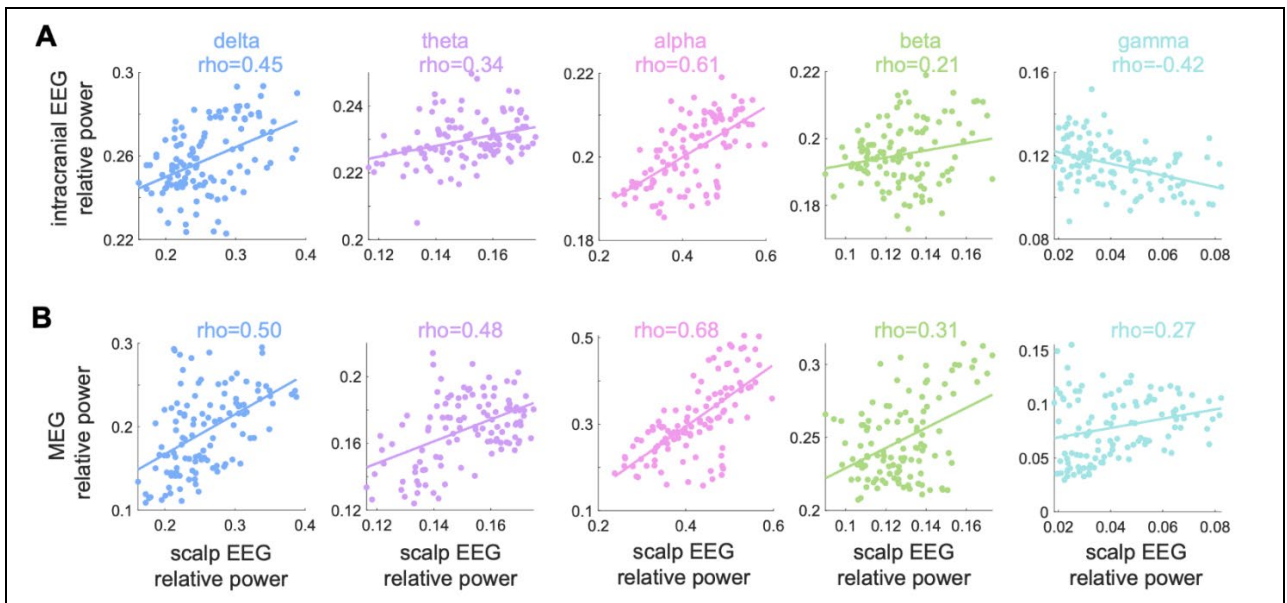

*Figure 3: Correlating scalp EEG to intracranial and MEG normative maps. Illustrating the similarity of the scalp EEG normative map with (A) intracranial EEG and (B) MEG normative maps across five individual band powers. The relative band powers are more positively correlated with scalp and MEG than scalp and intracranial across all frequency bands. Mean relative gamma band power is negatively correlated between intracranial and scalp maps (rho = -0.42).*

## Lateralisation of scalp EEG abnormality in temporal lobe epilepsy

We next investigated a potential application of the normative mapping approach to identify abnormality in a temporal lobe epilepsy cohort. We therefore focus our next analysis on the temporal lobe regions only. The patient cohort was normalised against the scalp normative map to extract abnormalities (maximum absolute z-scores) across 13 bilateral temporal regions, defined as per Lausanne parcellation scheme. Scalp EEG band power abnormality in the temporal lobe epilepsy cohort was found to be more ipsilateral to the epileptogenic cortex than contralateral across all but two ROI, and maximally so in the entorhinal region (t=2.6, Fig 4). An example patient from Fig 4b, is visually illustrated in Fig 4c showing the abnormality distribution, with the entorhinal cortex indicated by the blue arrow.

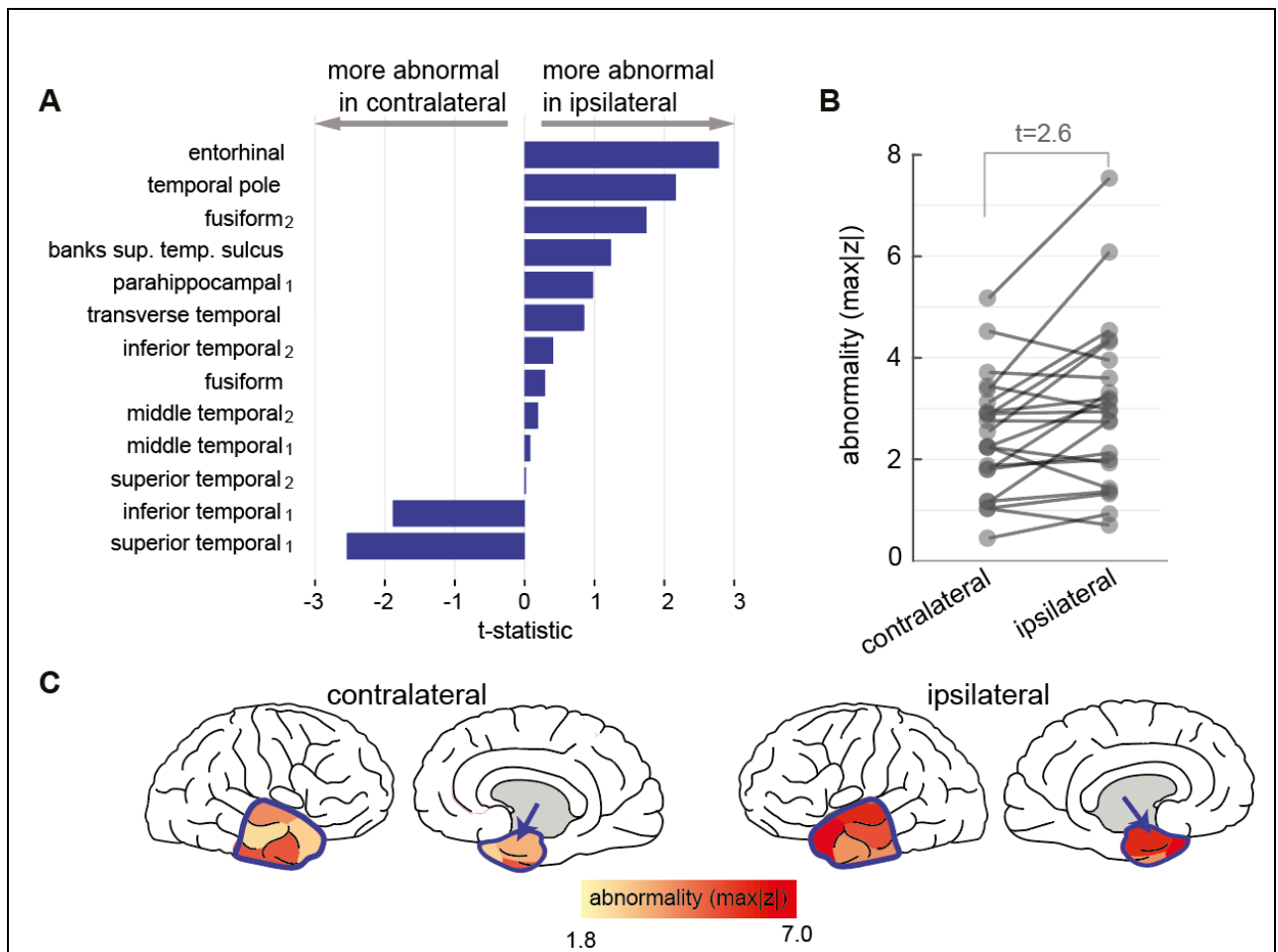

***Figure 4: Applying scalp EEG normative map to lateralise abnormality in temporal regions in individual patients.*** *(A) Paired T-test effect size of ipsilateral vs contralateral absolute z-score illustrating abnormality gradient across 13 bilateral temporal regions. (B) Entorhinal cortex data from panel a, shown as an absolute z-score in the ipsilateral and contralateral hemispheres of the 22 patients with a greater abnormality in the ipsilateral hemisphere (t=2.6). (C) Visualisation of max absolute z-score distribution in an example temporal epilepsy patient with score intensity indicating abnormality strength. The anterior temporal lobe is outlined by the blue contour, including entorhinal cortex (blue arrow), shows greater ipsilateral abnormality.*

## Discussion

In this study, we used scalp EEG relative band power to build a normative map across the cortex. Given the previous success of normative mapping using other modalities such as intracranial EEG and MEG, we broadened this approach by using a less expensive, and less invasive modality; scalp EEG. We present three main contributions. First, we built scalp EEG normative maps across 114 brain regions for five main frequency bands which we share openly with the scientific community, demonstrating inter-hemispheric symmetry and consistency of normative maps across time and duration. Second, we showed similarity of scalp EEG normative mapping across band power, with MEG and intracranial EEG. Third, we suggest a potentially useful clinical application by employing the normative map to lateralise regional abnormalities, across a patient cohort and within individual patients.

In our first analysis we found the temporal consistency of our normative map across epochs was extremely high for the three epochs. This is important to establish as high variability would render its subsequent use to localise abnormalities very challenging. Of note, however, is that our epochs were temporally proximal. For non-proximal epochs different brain states may occur. Specifically, it is well known that different brain states can affect EEG dynamics, for example daydreaming has been shown to enhance alpha power (Arnau et al., 2020; Compton et al., 2019), and the impact of circadian and ultradian rhythms remains a complex area of research (Lehnertz et al., 2021). Greater variations between states and duration may proportionally result in greater temporal differences, however this was not realisable in this study. Recordings of longer duration would be required to investigate such effects.

We next investigated inter-modality differences. The scalp normative map had strongly correlated spatial profiles with intracranial EEG and MEG (Frauscher et al., 2018; Keitel & Gross, 2016; Niso et al., 2019; Owen et al., 2022; Taylor et al., 2022). Relative alpha power was most spatially reproducible across all modalities, dominating the occipital and parietal regions in our work and in agreement with previous studies (Bosch-Bayard, Aubert-Vazquez, et al., 2020; Frauscher et al., 2018; Groppe et al., 2013; Nayak & Anilkumar, 2022; Owen et al., 2022; Taylor et al., 2022). High cross-modality similarity was also found in delta and theta band power profiles, reflecting strong signal in these bands. Normative maps were less similar across modalities for beta, and particularly gamma frequencies, and a negative correlation was found with intracranial EEG. High frequency activity originating from deep brain sources is hypothesised not to reach scalp surface and to be attenuated by the skull (Holler et al., 2018). Given that intracranial EEG electrodes are closer to the signal source, they are more sensitive to low amplitude fast frequencies such as gamma (Nottage & Horder, 2015), potentially explaining this finding. High inter-modality similarities across scalp, intracranial EEG and MEG propose future opportunities to explore multimodal analysis with addition of high-density scalp EEG.

In our exploratory clinical analysis we investigated the potential for lateralising abnormalities in a cohort of individuals with TLE who later had surgery to the hippocampus, amygdala and other temporal lobe areas. The amygdalohippocampal complex is one of the most important ictal generators as a common part of the epileptogenic network (Abel et al., 2018; Chabardes et al., 2005; Kahane & Bartolomei, 2010). However, as signal generated within deep tissue attenuates substantially before reaching cortical surface (Holler et al., 2018), amygdalae and hippocampi were excluded from this study. Despite this, the temporal pole was amongst the most asymmetrically abnormal regions. Other studies suggested a key role of the temporal pole in seizure networks (Abel et al., 2018; Chabardes et al., 2005; Kahane & Bartolomei, 2010). Alternative ictal generators including the parahippocampal and entorhinal cortex have been proposed (Bartolomei et al., 2004; Kahane & Bartolomei, 2010; Spencer & Spencer, 1994; Wennberg & Cheyne, 2014), and the fusiform gyrus in a single-patient case study (Oishi et al., 2002). Our results should be interpreted with caution however as a key limitation is that they are cohort based and not individualised. Therefore, for clinical translation an individual, personalised abnormality mapping may be useful to lateralise abnormalities.

This study has several strengths and limitations. One strength is the robustness of scalp normative maps across time and parcellations. Furthermore, similarity of our scalp normative maps with intracranial EEG and MEG modalities, also supported by the literature, collectively give confidence in our results. The limitations of this study include limited sample size from a single-site origin, narrowing the variability spectrum. The small number of scalp contacts limits signal precision. Furthermore, previous work suggests a potential age effect on band power profiles in children (Gomez et al., 2013; Gomez et al., 2017) and in elderly (Barry & De Blasio, 2017; Zhong & Chen, 2020), which was not investigated in our study of adult participants. We also used a single template MRI (ICMB152) to build realistic head models that limits the anatomical precision of source localisation.

Low-density scalp EEG has known spatial limitations in terms of localisation error, given the low spatial resolution (Burle et al, 2015) and poor conductivity of skin and other tissue (Shibata & Kobayashi, 2018). Furthermore, the skull imitating a low-pass filter further diminishes electrical potentials (Ramon et al. 2009; Srinivasan et al. 1998). There is view that interelectrode spacing of 3cm is a minimum requirement to avoid undersampling scalp potential, which requires approximately 100 electrodes (Kaiboriboon et al., 2012; Lantz et al., 2003; Spitzer et al., 1989). Previous studies confirm reduced localisation error with increased number of scalp electrodes (Lantz et al., 2003; Sohrabpour et al., 2015; Spitzer et al., 1989). There is also improved locasation with high density scalp electrodes (Holler et al., 2018) (Feyissa et al., 2017; Lantz et al., 2003). However, absolute improvement in localisation accuracy decreases with more electrodes (Sohrabpour et al., 2015). Ultimately, the ideal electrode configuration number is unknown. In this study, we used 30 scalp electrodes as this is commonly used in clinical settings (Brodbeck et al., 2010; Czigler et al., 2008; Kaiboriboon et al., 2012).

Many patients with focal epilepsy choose not to operate due to various reasons. Clinical utilisation of scalp EEG from a quantitate perspective may broaden diagnostic options, especially for difficult-to-treat patients by improving localisation. Further potential clinical application could be for other epilepsy syndromes and for predicting treatment outcomes (e.g. surgery, medication or stimulation), or even for diagnosis following first seizure. Taken together, this study presents a potential quantitative electrographic tool, which paves the way for further research of scalp EEG normative mapping and its possible clinical application in epilepsy and other neurological conditions.

# Acknowledgements

B.D. receives support from the NIH National Institute of Neurological Disorders and Stroke U01-NS090407 (Center for SUDEP Research) and Epilepsy Research UK. Y.W. is supported by a UKRI Future Leaders Fellowship (MR/V026569/1). P.N.T. is supported by a UKRI Future Leaders Fellowship (MR/T04294X/1). T.W.O is supported by the Centre for Doctoral Training in Cloud Computing for Big Data (EP/L015358/1). J.S.D. is grateful to Wellcome Trust (WT106882) and Epilepsy Research UK. We are grateful to the Epilepsy Society for supporting the Epilepsy Society MRI scanner. This work was supported by the National Institute for Health Research University College London Hospitals Biomedical Research Centre.

# References

Abel, T. J., Woodroffe, R. W., Nourski, K. V., Moritani, T., Capizzano, A. A., Kirby, P., Kawasaki, H., Howard, M., & Werz, M. A. (2018). Role of the temporal pole in temporal lobe epilepsy seizure networks: an intracranial electrode investigation. *J Neurosurg*, *129*(1), 165-173. https://doi.org/10.3171/2017.3.JNS162821

Arnau, S., Loffler, C., Rummel, J., Hagemann, D., Wascher, E., & Schubert, A. L. (2020). Inter-trial alpha power indicates mind wandering. *Psychophysiology*, *57*(6), e13581. https://doi.org/10.1111/psyp.13581

Arns, M., Conners, C. K., & Kraemer, H. C. (2013). A decade of EEG Theta/Beta Ratio Research in ADHD: a meta-analysis. *J Atten Disord*, *17*(5), 374-383. https://doi.org/10.1177/1087054712460087

Arya, R., Mangano, F. T., Horn, P. S., Holland, K. D., Rose, D. F., & Glauser, T. A. (2013). Adverse events related to extraoperative invasive EEG monitoring with subdural grid electrodes: a systematic review and meta-analysis. *Epilepsia*, *54*(5), 828-839. https://doi.org/10.1111/epi.12073

Barone, J., & Rossiter, H. E. (2021). Understanding the Role of Sensorimotor Beta Oscillations. *Front Syst Neurosci*, *15*, 655886. https://doi.org/10.3389/fnsys.2021.655886

Barry, R. J., & De Blasio, F. M. (2017). EEG differences between eyes-closed and eyes-open resting remain in healthy ageing. *Biol Psychol*, *129*, 293-304. https://doi.org/10.1016/j.biopsycho.2017.09.010

Bartolomei, F., Wendling, F., Regis, J., Gavaret, M., Guye, M., & Chauvel, P. (2004). Pre-ictal synchronicity in limbic networks of mesial temporal lobe epilepsy. *Epilepsy Res*, *61*(1-3), 89-104. https://doi.org/10.1016/j.eplepsyres.2004.06.006

Bernabei, J. M., Sinha, N., Arnold, T. C., Conrad, E., Ong, I., Pattnaik, A. R., Stein, J. M., Shinohara, R. T., Lucas, T. H., Bassett, D. S., Davis, K. A., & Litt, B. (2022). Normative intracranial EEG maps epileptogenic tissues in focal epilepsy. *Brain*, *145*(6), 1949-1961. https://doi.org/10.1093/brain/awab480

Blauwblomme, T., Ternier, J., Romero, C., Pier, K. S. T., D'Argenzio, L., Pressler, R., Cross, H., & Harkness, W. (2011). Adverse Events Occurring During Invasive Electroencephalogram Recordings in Children. *Operative Neurosurgery*, *69*, ons169-ons175. https://doi.org/10.1227/NEU.0b013e3182181e7d

Bonanni, L., Thomas, A., Tiraboschi, P., Perfetti, B., Varanese, S., & Onofrj, M. (2008). EEG comparisons in early Alzheimer's disease, dementia with Lewy bodies and Parkinson's disease with dementia patients with a 2-year follow-up. *Brain*, *131*(Pt 3), 690-705. https://doi.org/10.1093/brain/awm322

Bosch-Bayard, J., Aubert-Vazquez, E., Brown, S. T., Rogers, C., Kiar, G., Glatard, T., Scaria, L., Galan-Garcia, L., Bringas-Vega, M. L., Virues-Alba, T., Taheri, A., Das, S., Madjar, C., Mohaddes, Z., MacIntyre, L., Chbmp, Evans, A. C., & Valdes-Sosa, P. A. (2020). A Quantitative EEG Toolbox for the MNI Neuroinformatics Ecosystem: Normative SPM of EEG Source Spectra. *Front Neuroinform*, *14*, 33. https://doi.org/10.3389/fninf.2020.00033

Bosch-Bayard, J., Galan, L., Aubert Vazquez, E., Virues Alba, T., & Valdes-Sosa, P. A. (2020). Resting State Healthy EEG: The First Wave of the Cuban Normative Database. *Front Neurosci*, *14*, 555119. https://doi.org/10.3389/fnins.2020.555119

Brodbeck, V., Spinelli, L., Lascano, A. M., Pollo, C., Schaller, K., Vargas, M. I., Wissmeyer, M., Michel, C. M., & Seeck, M. (2010). Electrical source imaging for presurgical focus localization in epilepsy patients with normal MRI. *Epilepsia*, *51*(4), 583-591. https://doi.org/10.1111/j.1528-1167.2010.02521.x

Chabardes, S., Kahane, P., Minotti, L., Tassi, L., Grand, S., Hoffmann, D., & Benabid, A. L. (2005). The temporopolar cortex plays a pivotal role in temporal lobe seizures. *Brain*, *128*(Pt 8), 1818-1831. https://doi.org/10.1093/brain/awh512

Compton, R. J., Gearinger, D., & Wild, H. (2019). The wandering mind oscillates: EEG alpha power is enhanced during moments of mind-wandering. *Cogn Affect Behav Neurosci*, *19*(5), 1184-1191. https://doi.org/10.3758/s13415-019-00745-9

Czigler, B., Csikos, D., Hidasi, Z., Anna Gaal, Z., Csibri, E., Kiss, E., Salacz, P., & Molnar, M. (2008). Quantitative EEG in early Alzheimer's disease patients - power spectrum and complexity features. *Int J Psychophysiol*, *68*(1), 75-80. https://doi.org/10.1016/j.ijpsycho.2007.11.002

Duffy, F. H., Burchfiel, J. L., & Lombroso, C. T. (1979). Brain electrical activity mapping (BEAM): a method for extending the clinical utility of EEG and evoked potential data. *Ann Neurol*, *5*(4), 309-321. https://doi.org/10.1002/ana.410050402

Duffy, F. H., Iyer, V. G., & Surwillo, W. W. (1989). Clinical Use of Brain Electrical Activity Mapping. In *Clinical Electroencephalography and Topographic Brain Mapping: Technology and Practice* (pp. 222-237). Springer New York. https://doi.org/10.1007/978-1-4613-8826-5_19

Feyissa, A. M., Britton, J. W., Van Gompel, J., Lagerlund, T. L., So, E., Wong-Kisiel, L. C., Cascino, G. C., Brinkman, B. H., Nelson, C. L., Watson, R., & Worrell, G. A. (2017). High density scalp EEG in frontal lobe epilepsy. *Epilepsy Res*, *129*, 157-161. https://doi.org/10.1016/j.eplepsyres.2016.12.016

Fitzgerald, Z., Morita-Sherman, M., Hogue, O., Joseph, B., Alvim, M. K. M., Yasuda, C. L., Vegh, D., Nair, D., Burgess, R., Bingaman, W., Najm, I., Kattan, M. W., Blumcke, I., Worrell, G., Brinkmann, B. H., Cendes, F., & Jehi, L. (2021). Improving the prediction of epilepsy surgery outcomes using basic scalp EEG findings. *Epilepsia*, *62*(10), 2439-2450. https://doi.org/10.1111/epi.17024

Frauscher, B., von Ellenrieder, N., Zelmann, R., Dolezalova, I., Minotti, L., Olivier, A., Hall, J., Hoffmann, D., Nguyen, D. K., Kahane, P., Dubeau, F., & Gotman, J. (2018). Atlas of the normal intracranial electroencephalogram: neurophysiological awake activity in different cortical areas. *Brain*, *141*(4), 1130-1144. https://doi.org/10.1093/brain/awy035

Gomez, C., Perez-Macias, J. M., Poza, J., Fernandez, A., & Hornero, R. (2013). Spectral changes in spontaneous MEG activity across the lifespan. *J Neural Eng*, *10*(6), 066006. https://doi.org/10.1088/1741-2560/10/6/066006

Gomez, C. M., Rodriguez-Martinez, E. I., Fernandez, A., Maestu, F., Poza, J., & Gomez, C. (2017). Absolute Power Spectral Density Changes in the Magnetoencephalographic Activity During the Transition from Childhood to Adulthood. *Brain Topogr*, *30*(1), 87-97. https://doi.org/10.1007/s10548-016-0532-0

Groppe, D. M., Bickel, S., Keller, C. J., Jain, S. K., Hwang, S. T., Harden, C., & Mehta, A. D. (2013). Dominant frequencies of resting human brain activity as measured by the electrocorticogram. *Neuroimage*, *79*, 223-233. https://doi.org/10.1016/j.neuroimage.2013.04.044

Hagmann, P., Cammoun, L., Gigandet, X., Meuli, R., Honey, C. J., Wedeen, V. J., & Sporns, O. (2008). Mapping the structural core of human cerebral cortex. *PLoS Biol*, *6*(7), e159. https://doi.org/10.1371/journal.pbio.0060159

Hauser, W. A., Anderson, V. E., Loewenson, R. B., & McRoberts, S. M. (1982). Seizure Recurrence after a First Unprovoked Seizure. *New England Journal of Medicine*, *307*(9), 522-528. https://doi.org/10.1056/nejm198208263070903

He, X., Zhang, Y., Chen, J., Xie, C., Gan, R., Wang, L., & Wang, L. (2017). Changes in theta activities in the left posterior temporal region, left occipital region and right frontal region related to mild cognitive impairment in Parkinson's disease patients. *Int J Neurosci*, *127*(1), 66-72. https://doi.org/10.3109/00207454.2016.1143823

Holler, P., Trinka, E., & Holler, Y. (2018). High-Frequency Oscillations in the Scalp Electroencephalogram: Mission Impossible without Computational Intelligence. *Comput Intell Neurosci*, *2018*, 1638097. https://doi.org/10.1155/2018/1638097

Jensen, O., Goel, P., Kopell, N., Pohja, M., Hari, R., & Ermentrout, B. (2005). On the human sensorimotor-cortex beta rhythm: sources and modeling. *Neuroimage*, *26*(2), 347-355. https://doi.org/10.1016/j.neuroimage.2005.02.008

Kahane, P., & Bartolomei, F. (2010). Temporal lobe epilepsy and hippocampal sclerosis: lessons from depth EEG recordings. *Epilepsia*, *51 Suppl 1*, 59-62. https://doi.org/10.1111/j.1528-1167.2009.02448.x

Kaiboriboon, K., Luders, H. O., Hamaneh, M., Turnbull, J., & Lhatoo, S. D. (2012). EEG source imaging in epilepsy--practicalities and pitfalls. *Nat Rev Neurol*, *8*(9), 498-507. https://doi.org/10.1038/nrneurol.2012.150

Keitel, A., & Gross, J. (2016). Individual Human Brain Areas Can Be Identified from Their Characteristic Spectral Activation Fingerprints. *PLoS Biol*, *14*(6), e1002498. https://doi.org/10.1371/journal.pbio.1002498

Keizer, A. W. (2021). Standardization and Personalized Medicine Using Quantitative EEG in Clinical Settings. *Clin EEG Neurosci*, *52*(2), 82-89. https://doi.org/10.1177/1550059419874945

Klem, G. H., Lüders, H. O., Jasper, H. H., & Elger, C. (1999). The ten-twenty electrode system of the International Federation. The International Federation of Clinical Neurophysiology. *Electroencephalogr Clin Neurophysiol Suppl*, *52*, 3-6.

Ko, J., Park, U., Kim, D., & Kang, S. W. (2021). Quantitative Electroencephalogram Standardization: A Sex- and Age-Differentiated Normative Database. *Frontiers in Neuroscience*, *15*. https://doi.org/10.3389/fnins.2021.766781

Lantz, G., Grave de Peralta, R., Spinelli, L., Seeck, M., & Michel, C. M. (2003). Epileptic source localization with high density EEG: how many electrodes are needed? *Clin Neurophysiol*, *114*(1), 63-69. https://doi.org/10.1016/s1388-2457(02)00337-1

Lehnertz, K., Rings, T., & Bröhl, T. (2021). Time in Brain: How Biological Rhythms Impact on EEG Signals and on EEG-Derived Brain Networks. *Frontiers in Network Physiology*, *1*. https://doi.org/10.3389/fnetp.2021.755016

Livinț Popa, L., Dragoș, H. M., Strilciuc, Ș., Pantelemon, C., Mureșanu, I., Dina, C., Văcăraș, V., & Mureșanu, D. (2021). Added Value of QEEG for the Differential Diagnosis of Common Forms of Dementia. *Clin EEG Neurosci*, *52*(3), 201-210. https://doi.org/10.1177/1550059420971122

Michel, C. M., & Brunet, D. (2019). EEG Source Imaging: A Practical Review of the Analysis Steps. *Front Neurol*, *10*, 325. https://doi.org/10.3389/fneur.2019.00325

Morillon, B., Arnal, L. H., Schroeder, C. E., & Keitel, A. (2019). Prominence of delta oscillatory rhythms in the motor cortex and their relevance for auditory and speech perception. *Neurosci Biobehav Rev*, *107*, 136-142. https://doi.org/10.1016/j.neubiorev.2019.09.012

Nayak, C. S., & Anilkumar, A. C. (2022). EEG Normal Waveforms. In *StatPearls*. StatPearls Publishing Copyright © 2022, StatPearls Publishing LLC. https://www.ncbi.nlm.nih.gov/books/NBK539805/

Niso, G., Tadel, F., Bock, E., Cousineau, M., Santos, A., & Baillet, S. (2019). Brainstorm Pipeline Analysis of Resting-State Data From the Open MEG Archive. *Frontiers in Neuroscience*, *13*. https://doi.org/10.3389/fnins.2019.00284

Nottage, J. F., & Horder, J. (2015). State-of-the-Art Analysis of High-Frequency (Gamma Range) Electroencephalography in Humans. *Neuropsychobiology*, *72*(3-4), 219-228. https://doi.org/10.1159/000382023

Oishi, M., Kameyama, S., Morota, N., Tomikawa, M., Wachi, M., Kakita, A., Takahashi, H., & Tanaka, R. (2002). Fusiform gyrus epilepsy: the use of ictal magnetoencephalography: Case report. *Journal of Neurosurgery*, *97*(1), 200-204. https://doi.org/10.3171/jns.2002.97.1.0200

Owen, T. W., Schroeder, G. M., V., J., Hall, G. R., McEvoy, A., Miserocchi, A., de Tisi, J., Duncan, J. S., Rugg-Gunn, F., Wang, Y., & Taylor, P. N. (2022). MEG abnormalities highlight mechanisms of surgical failure in neocortical epilepsy. 33.

Ramon, C., Freeman, W. J., Holmes, M., Ishimaru, A., Haueisen, J., Schimpf, P. H., & Rezvanian, E. (2009). Similarities between simulated spatial spectra of scalp EEG, MEG and structural MRI. *Brain Topogr*, *22*(3), 191-196. https://doi.org/10.1007/s10548-009-0104-7

Shibata, T., & Katsuhiro, K. (2018). Epileptic High-frequency Oscillations in Scalp Electroencephalography. *Acta Med. Okayama*, *72*(4), 325-329.

Snipes, S., Krugliakova, E., Meier, E., & Huber, R. (2022). The Theta Paradox: 4-8 Hz EEG Oscillations Reflect Both Sleep Pressure and Cognitive Control. *J Neurosci*, *42*(45), 8569-8586. https://doi.org/10.1523/JNEUROSCI.1063-22.2022

Sohrabpour, A., Lu, Y., Kankirawatana, P., Blount, J., Kim, H., & He, B. (2015). Effect of EEG electrode number on epileptic source localization in pediatric patients. *Clin Neurophysiol*, *126*(3), 472-480. https://doi.org/10.1016/j.clinph.2014.05.038

Spencer, S. S., & Spencer, D. D. (1994). Entorhinal-hippocampal interactions in medial temporal lobe epilepsy. *Epilepsia*, *35*(4), 721-727. https://doi.org/10.1111/j.1528-1157.1994.tb02502.x

Spitzer, R. A., Cohen, L. G., Fabrikant, J., & Hallett, M. (1989). A method for determining optimal interelectrode spacing for cerebral topographic mapping. 355-361.

Srinivasan, R., Tucker, M. D., & Murias, M. (1998). Estimating the spatial Nyquist of the human EEG. *Behavior Research Methods, Instruments, & Computers*, *30*(1), 8-19.

Sun, J., Tang, Y., Lim, K. O., Wang, J., Tong, S., Li, H., & He, B. (2014). Abnormal dynamics of EEG oscillations in schizophrenia patients on multiple time scales. *IEEE Trans Biomed Eng*, *61*(6), 1756-1764. https://doi.org/10.1109/TBME.2014.2306424

Tadel, F., Baillet, S., Mosher, J. C., Pantazis, D., & Leahy, R. M. (2011). Brainstorm: a user-friendly application for MEG/EEG analysis. *Comput Intell Neurosci*, *2011*, 879716. https://doi.org/10.1155/2011/879716

Taylor, P. N., Papasavvas, C. A., Owen, T. W., Schroeder, G. M., Hutchings, F. E., Chowdhury, F. A., Diehl, B., Duncan, J. S., McEvoy, A. W., Miserocchi, A., de Tisi, J., Vos, S. B., Walker, M. C., & Wang, Y. (2022). Normative brain mapping of interictal intracranial EEG to localize epileptogenic tissue. *Brain*, *145*(3), 939-949. https://doi.org/10.1093/brain/awab380

Tedrus, G. M., Negreiros, L. M., Ballarim, R. S., Marques, T. A., & Fonseca, L. C. (2019). Correlations Between Cognitive Aspects and Quantitative EEG in Adults With Epilepsy. *Clin EEG Neurosci*, *50*(5), 348-353. https://doi.org/10.1177/1550059418793553

Valdes-Sosa, P. A., Galan-Garcia, L., Bosch-Bayard, J., Bringas-Vega, M. L., Aubert-Vazquez, E., Rodriguez-Gil, I., Das, S., Madjar, C., Virues-Alba, T., Mohades, Z., MacIntyre, L. C., Rogers, C., Brown, S., Valdes-Urrutia, L., Evans, A. C., & Valdes-Sosa, M. J. (2021). The Cuban Human Brain Mapping Project, a young and middle age population-based EEG, MRI, and cognition dataset. *Sci Data*, *8*(1), 45. https://doi.org/10.1038/s41597-021-00829-7

Varatharajah, Y., Berry, B., Joseph, B., Balzekas, I., Pal Attia, T., Kremen, V., Brinkmann, B., Iyer, R., & Worrell, G. (2021). Characterizing the electrophysiological abnormalities in visually reviewed normal EEGs of drug-resistant focal epilepsy patients. *Brain Commun*, *3*(2), fcab102. https://doi.org/10.1093/braincomms/fcab102

Wennberg, R., & Cheyne, D. (2014). EEG source imaging of anterior temporal lobe spikes: validity and reliability. *Clin Neurophysiol*, *125*(5), 886-902. https://doi.org/10.1016/j.clinph.2013.09.042

Zhong, X., & Chen, J. J. (2020). Variations in the frequency and amplitude of resting-state EEG and fMRI signals in normal adults: The effects of age and sex. *bioRxiv*. https://doi.org/10.1101/2020.10.02.323840

# Supplementary

**Supplementary Material 1: Additional information on individual subjects and their EEG**

In the main manuscript we provided descriptive statistics of patients and healthy controls in terms of age, sex (M/F), side (Left/Right), age of onset and duration in Table 1. Here we provide more information on individual subject basis and channels that were removed for individual subject during pre-processing.

**Table S1: Additional details of patient data**

| Patient ID | Age | Age at Onset | Duration | Sex | Side | First Pathology | Channels Removed |
|---|---|---|---|---|---|---|---|
| 1 | 24 | 15 | 9 | F | Left | Other | - |
| 2 | 29 | 15 | 14 | F | Left | HS | - |
| 3 | 31 | 13 | 18 | F | Left | HS (EFS) | - |
| 4 | 31 | 13 | 18 | F | Left | HS (EFS) | - |
| 5 | 33 | 15 | 18 | M | Right | Other | - |
| 6 | 19 | 7.5 | 11.5 | F | Left | HS | - |
| 7 | 42 | 3 | 39 | F | Left | Other | TP10, O2 |
| 8 | 24 | 19 | 5 | M | Right | Other | - |
| 9 | 47 | 14 | 33 | M | Right | Other | - |
| 10 | 25 | 9 | 16 | F | Left | HS | F7, F8 |
| 11 | 38 | 30 | 8 | F | Right | HS | - |
| 12 | 48 | 4 | 44 | M | Left | HS | - |
| 13 | 47 | 9 | 38 | F | Left | HS | - |
| 14 | 57 | 5 | 52 | M | Right | Other | O1, O2, Oz |
| 15 | 33 | 2 | 31 | M | Left | HS | - |
| 16 | 26 | 7 | 19 | M | Left | HS | - |
| 17 | 35 | 18 | 17 | M | Right | Other | TP9 |
| 18 | 31 | 24 | 7 | F | Left | Other | - |
| 19 | 32 | 10 | 22 | M | Left | HS | Fp1, Fp2, F8 |
| 20 | 45 | 17 | 28 | F | Right | HS (EFS) | P7, C3 |
| 21 | 19 | 3 | 16 | F | Left | HS (EFS) | - |
| 22 | 37 | 32 | 5 | F | Right | Other | Fp2 |

**Table S2: Details of healthy control data**

| Control ID | Age | Sex | Channels Removed |
|---|---|---|---|
| 1 | 28 | F | - |
| 2 | 39 | M | - |
| 3 | 49 | F | - |
| 4 | 29 | F | - |
| 5 | 41 | M | - |
| 6 | 35 | M | - |
| 7 | 28 | M | - |
| 8 | 30 | M | - |
| 9 | 30 | F | - |
| 10 | 33 | F | - |
| 11 | 28 | M | T7 |
| 12 | 22 | M | - |
| 13 | 30 | M | - |
| 14 | 28 | M | T8 |
| 15 | 34 | M | - |
| 16 | 25 | F | T8 |
| 17 | 33 | M | - |

**Supplementary Material 2: Estimating normative map robustness – Epoch 3**
In the main manuscript we presented correlation and standard deviation graphs of normative map 1 vs map 2. Here we present additional figures that involve map 3 and their respective correlation results

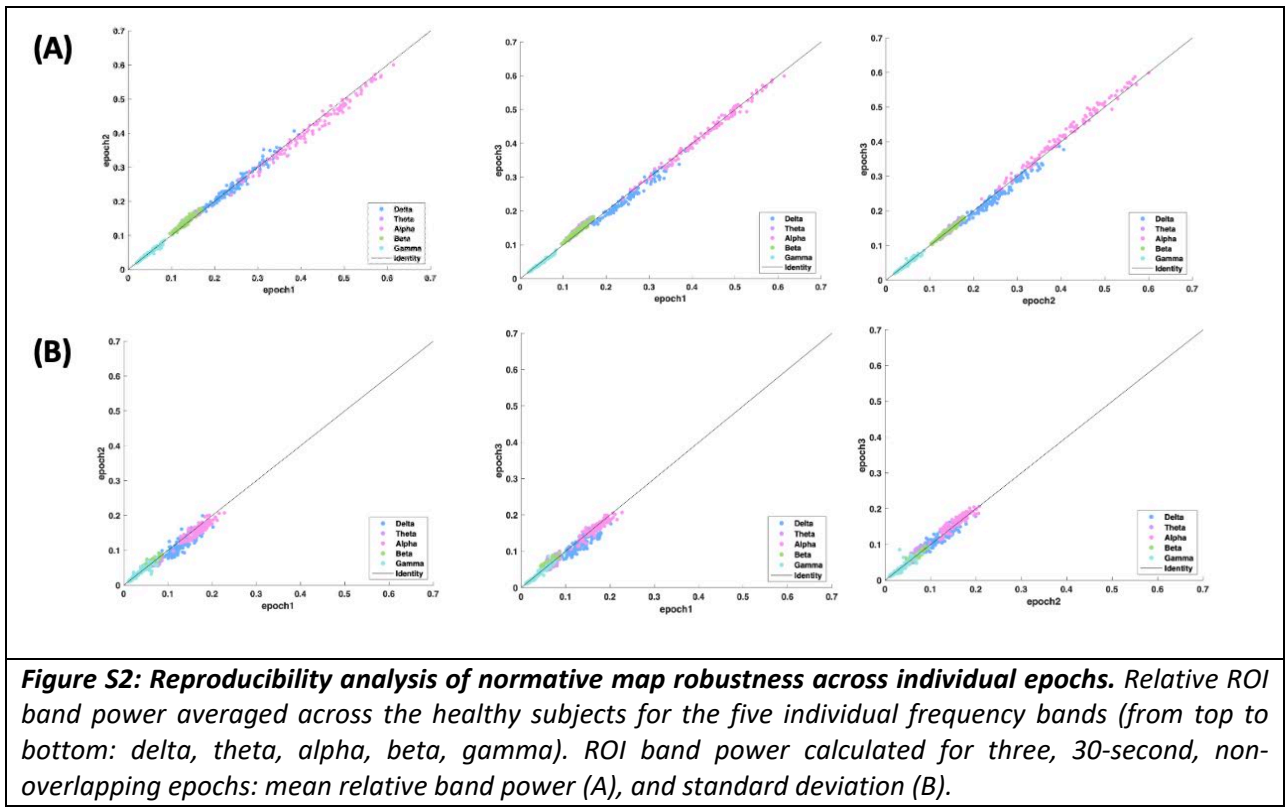

*Figure S2: Reproducibility analysis of normative map robustness across individual epochs. Relative ROI band power averaged across the healthy subjects for the five individual frequency bands (from top to bottom: delta, theta, alpha, beta, gamma). ROI band power calculated for three, 30-second, non-overlapping epochs: mean relative band power (A), and standard deviation (B).*

**Supplementary Material 3: Normative map consistency across parcellations**

In the main text we adapted the Lausanne parcellation of scale 60 with 114 regions of interest. Here we illustrate similarity Comparison of scalp normative maps across three alternative parcellations. Resolution sizes of the following include: 68, 219 and 448 regions, respectively.

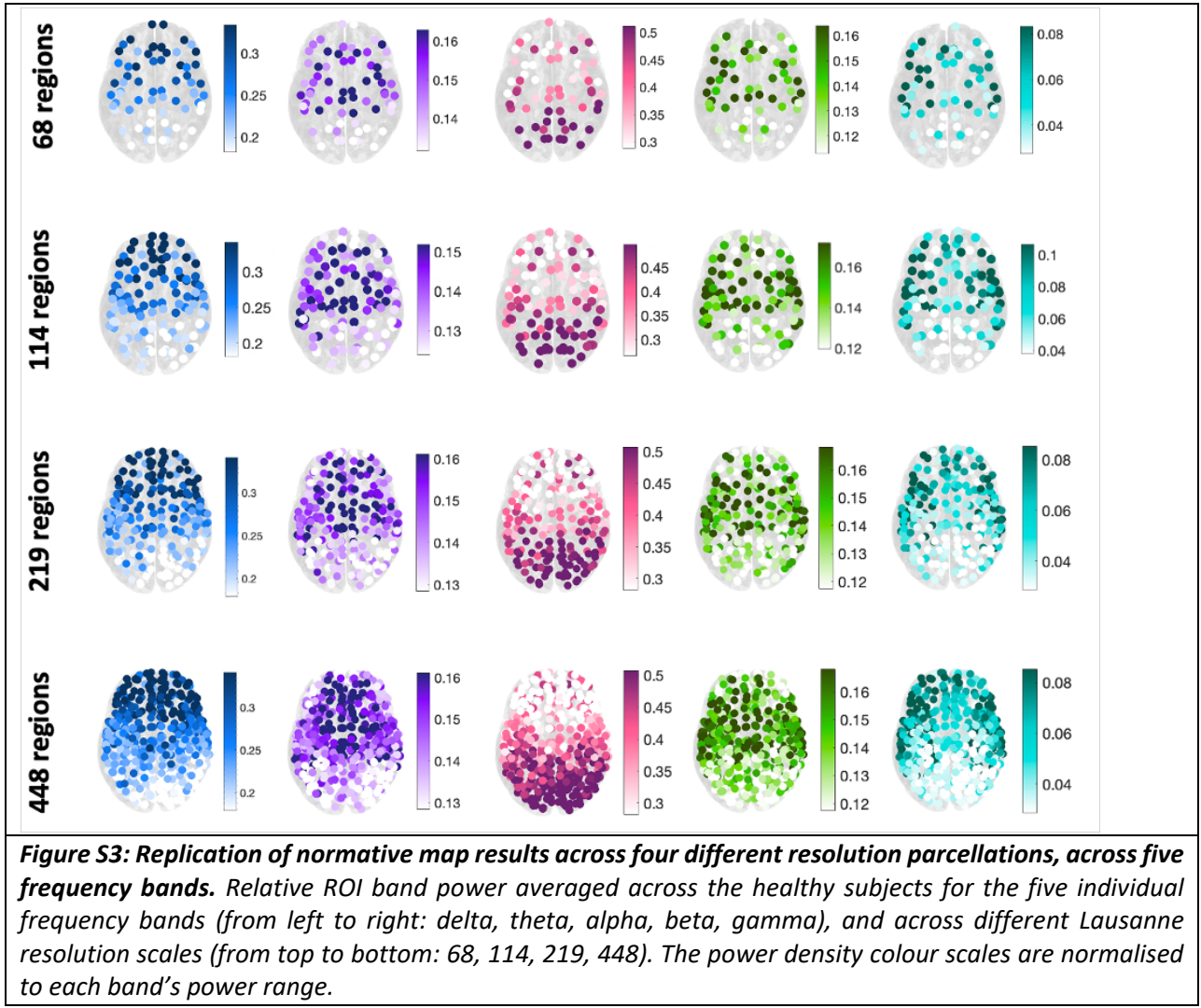

***Figure S3: Replication of normative map results across four different resolution parcellations, across five frequency bands.*** *Relative ROI band power averaged across the healthy subjects for the five individual frequency bands (from left to right: delta, theta, alpha, beta, gamma), and across different Lausanne resolution scales (from top to bottom: 68, 114, 219, 448). The power density colour scales are normalised to each band's power range.*